\documentclass{elsart}
\usepackage{latexsym}
\usepackage[final]{epsfig}
\newcommand{\avg}[1]{\langle{#1}\rangle}
\newcommand{\be}{\begin{equation}}
\newcommand{\ee}{\end{equation}}
\newcommand{\beas}{\begin{eqnarray*}}
\newcommand{\eeas}{\end{eqnarray*}}
\newcommand{\bea}{\begin{eqnarray}}
\newcommand{\eea}{\end{eqnarray}}
\newcommand{\req}[1]{(\ref{#1})}
\newcommand{\acca}{h^{\alpha}}

\begin{document}
\begin{frontmatter}
\title{Buyer feedback as a filtering mechanism for reputable sellers}
%
\author[Frib]{Paolo Laureti\thanksref{ePL}},
\author[Prague]{Franti\v{s}ek Slanina},
\author[Flo]{Yi-Kuo Yu} and
\author[Frib]{Yi-Cheng Zhang}
\address[Frib]{
        Institut de Physique Th\'eorique, Universit\'e
        de Fribourg,
        P\'erolles, CH-1700 Fribourg, Switzerland
}
\address[Prague]{
        Institute of Physics,
        Academy of Sciences of the Czech Republic,
        Na~Slovance~2, CZ-18221~Praha,
        Czech Republic
}
\address[Flo]{
        Department of Physics,
        Florida Atlantic University,
        Boca Raton, FL 33431,
        USA
}
\thanks[ePL]{paolo.laureti@unifr.ch}
%
\date{\today}
\begin{abstract}

We propose a continuum model for the description of buyer and seller 
dynamics in an Internet market.
The relevant variables are the research effort of buyers and the sellers' 
reputation building process.
We show that, if a commercial web-site gives consumers the possibility to 
rate credibly sellers they bargained with,
vendors are forced to be more honest. This leads to mutual beneficial 
symbiosis between buyers and sellers; the
overall enhanced volume of transactions contributes ultimately to the web-site, which 
facilitates the matchmaking service.
\end{abstract}

\begin{keyword}Asymmetric Information; Internet Commerce;
Game theory; Self-organization; Symbiosis.
\\
{\it PACS: }
05.65.+b; 
02.50.Le; 
87.23.Ge 
\end{keyword}
\end{frontmatter}
\section{The Problem}
The Internet provides a new venue for commercial transactions, though there 
is still no consensus as to what fundamental
mechanism makes a commercial web site tick or flop. In the law of supply and demand, 
transactions are beneficial to both buyers and
sellers in general. However, so-called market failures can happen if the 
information about the quality of the product is very
asymmetric. In the last few decades economists have made fundamental 
research work in this area ---for instance the famous paper by
George Akerlof on the "Lemon's Problem" \cite{akerlof} as well as more general 
applications of asymmetric information in many
economic relationships by Joe Stiglitz et al.; for a recent review see \cite{Stiglitz}.

What is special about Internet commerce? In the last few years the much 
initial enthusiasm
turned into big disappointment, after many high flyers crashed and the 
Internet commerce bubble blowed. We feel that the
fundamental mechanism is not yet generally appreciate. In this paper we 
want to highlight the unique role played by the
reputation system. In the Internet commerce, the information asymmetry is 
extreme: buyers cannot evaluate the quality of products
before purchasing them. Even worse, buyers don't even see sellers in their 
face, as in an off-line transaction. Thus the
information asymmetry is much more severe than in the traditional commerce 
modes. The Internet, on the other hand, offers
tremendous opportunities, since buyers can access a vast choice of 
products and the search costs are much reduced.

So, we face the dilemma: how to tap into
the huge potential while avoiding the proverbial information asymmetry? Our analysis will show that the holy grail rests in
binding the collective knowledge of all buyers about the sellers' reputation. The Internet commerce has 
the unprecedented potential to leverage the collective buying experience in a centralized place. Though a less than 
honest seller can get away with a questionable transaction on one buyer, the dissatisfied buyer can easily post his rating 
on this particular seller. Reputation is a valuable asset that no vendor can ignore. 
Indeed, most of the fast 
growing e-commerce web sites, including Internet auctions sites such as 
eBay \cite{auction}, allow buyers to rate sellers after receiving the product they bid for. 
The ensemble of those ratings builds up a seller's reputation that can be viewed ever since by other buyers, thus 
replacing (and sometimes improving) the direct quality check of usual street 
shopping.

Our analysis is based on the fundamental conviction that 
sellers have the option of being honest or not. If the web site can establish a credible rating system to capture buyers' feedback, it 
effectively filters out dishonest sellers. There are no permanent cheaters, they must find other ways to make a living that 
are also beneficial to the society. The matchmaking service, provided by the web-site, facilitates a selection process that is the opposite of "adverse selection" \cite{Arrow}. Those sellers with good quality, would be encourage to join by a honest representation, 
while the dishonest ones wouldn't even want to try. Such a service can be easily rewarded, since such a web-site can take a slice 
from the mutually beneficial transactions. Our results show that the extent of the total transactions depends 
on the quality of the feedback rating system. 

Our approach is to model buyers and sellers as two species in symbiosis, much like in population dynamics. In fact, 
our equations share many similarities with the well-known Lotka-Volterra model \cite{Lotka,volterra}. The key is to realize the two species have 
both converging as well as diverging interests: Buyers need sellers, the more the better, and vice versa. This much they have
converging interests. But, in a particular transaction, a buyer's loss is the seller's gain. However, on the aggregate level 
transactions between the two groups are non-zero sum games ---in fact positive sum games. 
 
\section{The Model}
At time $t$, $B(t)$ potential buyers and $S(t)$ sellers meet in our virtual marketplace.
Buyers' rationality is bounded by 
incomplete information and limited computing capability, but we retain the 
assumption of procedural rationality \cite{simon}. This means agents only 
dispose of a few options, but they are able to choose better ones with 
higher probability, provided they are given correct information. We shall 
consider a fast growing regime, assuming the number of buyers $B(t)$ grows 
exponentially in time. Regular users may decide to continue trading or stop 
doing so, according to their satisfaction at previous times.

The interest a given seller has in staying on that market can be easily 
estimated as a function of the earnings made. If he gained nothing, it is 
very probable that he won't repeat the experience. Sometimes he may lower 
his honesty in the hope to make more money, or increase it in order to sell 
more.
On the other hand, buyers' profitability has to be inferred. Neglecting 
possible technical dysfunctions, delivery failures and other inconvenients 
not directly related to the actors of the transaction, there are two main 
possible sources of discontent: the item could differ from what the buyer 
was originally looking for, or its state of usage could be worse than he was 
promised. If a buyer is not satisfied, he is not likely to visit the 
web-site again in the near future.

Each seller $s$ only sells products of a kind ($x_s$) and is characterized 
by his honesty $h_s$,
which can be interpreted as the ratio quality/price he is selling at.
His satisfaction $\gamma_s(t)$ is defined as the number of products sold 
$n_s(t)$, times the normalized unitary gain $g_{h_s}$:
\bea
\gamma_s(t)&=&n_s(t) g_{h_s}\\\label{gain}
g_{h_s}&=&1-h_s+m,
\eea
where $m$ is the minimum profit margin, i.e. a percentage of the price that 
covers all expenses and leaves a revenue even to the most honest vendors. 
Here $1-h_s$ can be regarded as an extra-profit that would equal zero in a 
perfectly competitive market. We assume no price discrimination on an 
individual basis, i.e. $h_s$ does not depend on the particular buyer $s$ is 
dealing with, even though this phenomenon may arise in particular contests 
\cite{varian}.

Each buyer $b$ looks, at time $t$, for a specifically desired product $x_b$.
Products $x_k$, be they desired or sold, can be represented as elements of a 
metric space (real numbers or bit strings), where we can define a normalized 
distance $d_{b,s}=d(x_b,x_s)\in[0,1)$ and an overlap $q_{b,s}=1-d_{b,s}$.
The latter measures how close a product $x_s$ is to the buyer's desire 
$x_b$. Now, if $b$ buys a unit of $x_{\tilde s}$, once he receives it he is 
rewarded with a payoff $r_{b,{\tilde s}}$. His satisfaction $\gamma_b(t)$ 
then equals:
\be
\gamma_b(t)=\left\{
\begin{array}{ll}
r_{b,\tilde{s}} & \textrm{if $b$ purchased $x_{\tilde s}$ at time $t$;}\\
0             & \textrm{if $b$ purchased nothing at time $t$;}
\end{array}\right.
\ee
It is reasonable that $r_{b,s}$ be an increasing function of $h_s$ (buyers 
are more satisfied if the purchased product has a better ratio 
quality/price) and $q_{b,s}$ (buyers are more satisfied if the purchased 
product is closer to their wishes). Hence we define
\be\label{rew}
r_{b,s}=h_s q_{b,s}.
\ee
Finally, buyer $b$ can rate seller's $s$ honesty and influence his 
reputation. This happens for every buyer who deals with seller $s$ at each 
time step, therefore one's reputation tends to his honesty $h_s$. Buyers, 
then, can take a look at sellers' reputation before purchasing a product. 
Whenever a difference between reputation and honesty is not explicitly 
mentioned, we shall assume they coincide. On the other hand, buyers are 
allowed to trust it or not, in a way we will describe later on.

Now we have a definition of buyers' and sellers' satisfaction. Their role 
becomes clear once we specify the dynamics: we will do that first, leaving 
the details of the transaction process for later sections. Since we are 
aiming to give a mean field description of the system, it is useful to 
introduce the average buyers' and sellers' satisfactions
\bea\label{1gamma}
\Gamma_B(t)=\sum_b\gamma_b(t)/B(t)\\\label{2gamma}
\Gamma_S(t)=\sum_s\gamma_s(t)/S(t).
\eea
When the number of buyers and sellers becomes large, the self-averaging 
effect yields $\sum_{\zeta}\gamma_{\zeta}(t)\simeq\sum_{\zeta}{\bar 
\gamma}_{\zeta}(t)$, with $\zeta=b,s$. Here the overline bar is an average 
over realizations: ${\bar \gamma}_{\zeta}(t)$ represents the average payoff 
an agent would get if he faced the same situation a great number of times.
Let us assume $h$ is a discrete variable which can only take $H$ values, 
separated by a mesh $\Delta h=1/(H-1)$. Then we can consider the number of 
sellers $S_h(t)$ belonging to a certain honesty class, with 
$S(t)=\sum_{h=0}^1 S_h(t)$. Their average satisfaction will be
\bea
\Gamma_{S_h}(t)&=&\sum_{s: h_s=h}\gamma_s(t)/S_h(t), \textrm{  with:}\\ 
\label{gaso}
\Gamma_S(t)&=&\sum_h S_h(t) \Gamma_{S_h}(t)/S(t)=\avg{\Gamma_{S_h}(t)}.
\eea
Here and in the following, angular brackets stand for averages over the 
honesty distribution
\be\label{pidiac}
p(h,t)=S_h(t)/S(t).
\ee
Notice that, while $\Gamma_B(t)$ is constrained in the range $[0,1]$, the 
value of $\Gamma_{S_h}$ is only bounded by $(m+1)B(t)$.
Now we are able to write $H+1$ replicator dynamics type \cite{replica} 
differential equations, describing the mean field time evolution of $B(t)$ 
and $S_h(t)$, for $h=0, \Delta h, 2\Delta h, ..., 1$:
\bea\label{buydyn}
\frac{dB(t)}{dt}&=&c_B B(t) - [1-\Gamma_B(t)]B(t)\\\label{seldyn}
\frac{dS_h(t)}{dt}&=& \Delta h [\Gamma_S(t)-1] S(t)+ 
[\Gamma_{S_h}(t)-1]S_h(t).
\eea
Here the parameter $c_B$ is a factor of growth, which embeds all the 
external conditions, such as liquidity and competition effects.
Summing equation \req{seldyn} over $h$ we obtain
\be\label{seldyntot}
\frac{dS(t)}{dt}= 2[\Gamma_S(t)-1] S(t).
\ee

The above equations arise from the following dynamics. At time $t$ every 
buyer attracts $c_B$ users in the web-site. Among the old clients a 
percentage $\Gamma_B\in[0,1]$ survives, while the others leave. In other 
words, the probability that an active buyer continues shopping in this 
market at future times is proportional to his satisfaction at time $t$.
As for sellers, the term $\Delta h [\Gamma_S(t)-1] S(t)$ on the r.h.s. of 
equation \req{seldyn} acts uniformly on every $h$ level. Since $\Gamma_S(t)$ is 
the average profit a seller made at time $t$, it represents a general 
measure of profitability for the web-site. If it is bigger than a given 
value, which we arbitrarily posit equal to one, new sellers are likely to 
add listings to the web-site.
We can think they are people who only look at aggregate results before 
entering a market, non professional vendors drawn from a uniform honesty 
distribution. If $\Gamma_S<1$ some of these persons will drop out, with the 
understanding that $S_h(t)$ be set to zero if it falls below it.
The second term $[\Gamma_{S_h}(t)-1]S_h(t)$ of equation \req{seldyn} is 
strongly $h$-dependent.
When it is smaller than one, a percentage $1-\Gamma_{S_h}$ of sellers of 
honesty $h$ drops out, vice-versa when $\Gamma_{S_h}>1$.
In this case the newcomers are fairly well informed about the market 
dynamics and  estimate how much extra-profit they can make, thus choosing a 
specific entry honesty level. Notice that the honesty $h_s$ of a given 
active seller cannot be changed in time, but $s$ can always exit and come 
back with a more profitable one.

The functions $\Gamma$ depend on the probability distribution $\mu(q)$ of 
the overlap, arising from the choice of the metric space of products, and on 
the amount of information buyers collect before purchasing an item. In the 
following sections we shall analyze two particular cases. First we shall 
model consumers going for one specific product (maximal selection); then 
flexible ones, looking for a product similar ``enough'' to their wishes 
(browsing agents).

\section{Maximal selection}
Here we analyze a process where potential consumers decide whether to buy or 
not a single particular item per time unit.
As we already mentioned, a buyer $b$ access the web-site looking for a 
desired product $x_b$. Now he considers what is available in the market, 
picks the item that fits best his request, decides whether to buy it or not, 
and finally he may receive and judge it.
Let us assume that, thanks to internal search tools of the web-site, he 
finds the item $x_{\hat s_b}$ corresponding to the maximum overlap 
$q_{b,{\hat s_b}}=\max_{s}q_{b,s}$. Then he evaluates it, checking the 
seller's reputation, and decides if he wants to buy it or not, with no 
further research. He purchases it with probability $f_b({\hat s_b})$, 
proportional to the buyer's expected reward.
The latter can differ from the actual payoff $r_{i,{\hat s_b}}$ (\ref{rew}) 
he would eventually get from the purchase. In fact, at this stage, the buyer 
does not have the product $x_{\hat s_b}$ in his hands and can only guess 
upon the available information. He could, therefore, trust differently his 
perception of $h_s$ and $q_{b,s}$, the first one coming from other buyers' 
ratings of seller $\hat{s_b}$, the second from a description (sometimes a 
picture) of the item, provided by the seller himself. Hence we define
\be\label{pur}
f_b(s)=\acca_s q_{b,s},
\ee
where the exponent $\alpha$ is a parameter that tunes the weight consumers 
give to sellers' reputation. If he decides to buy, $b$ eventually receives 
the product, rates seller ${\hat s_b}$ with $h_{\hat s_b}$ and is rewarded 
with a payoff $r_{b,{\hat s_b}}$. The average satisfaction then equals
\be
\Gamma_B(t)=\frac{1}{B(t)}\sum_b f_b({\hat s_b})r_{b,{\hat 
s_b}}=\frac{1}{B(t)}\sum_b h_{\hat s_b}^{\alpha+1}q_{b,{\hat s_b}}^2.
\ee
When we take the average over all buyers, we are implicitly averaging over 
the honesty distribution $p(h,t)$ \req{pidiac}, because the index ${\hat s_b}$ 
depends on the chosen seller. Let us approximate $q_{b,{\hat s_b}}$ with its 
average value over all buyers $q_{max}$; then
\be\label{gammabm}
\Gamma_B(t)=\avg{h^{\alpha+1}} q_{max}^2.
\ee

\begin{figure}
\epsfig{file=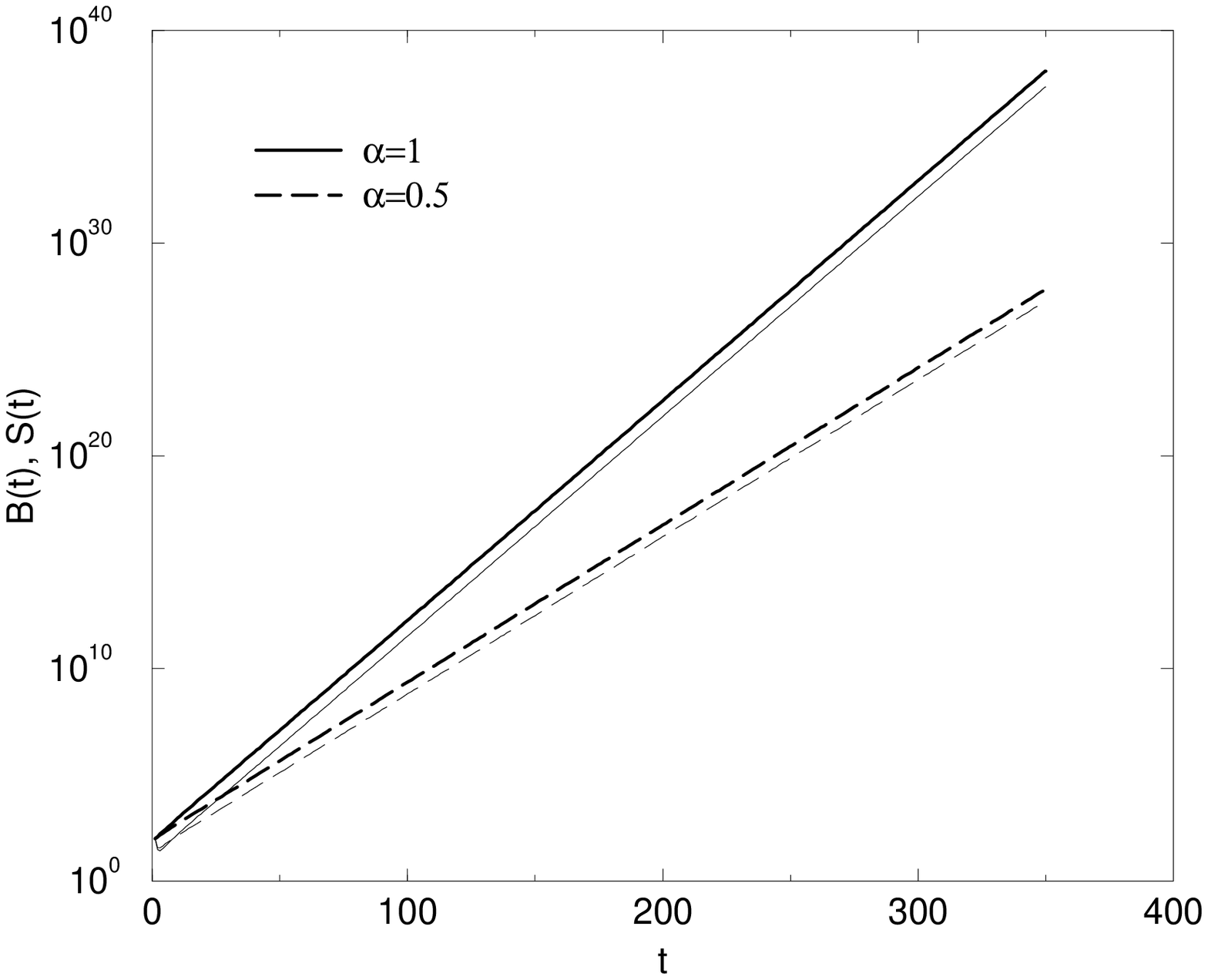, width=250pt}
\caption{Buyers (bold lines) and sellers growth as a function of time, with 
maximal selection. } \label{fig2}
\end{figure}

Every seller has equal probability to maximize the overlap of a given buyer. 
Conversely their probability to sell a product once chosen, and their 
unitary profit, depend on their honesty level. The average profit made by a 
seller of honesty $h$ then reads
\bea\nonumber
\Gamma_{S_h}(t)&=&N_h(t) g_h\\ \label{gammashm}
&=&\acca q_{max}\frac{B(t)}{S(t)}(1-h+m),
\eea
where $N_h(t)=\sum_{s: h_s=h} n_s(t)/S_h(t)$ is the average number of 
items sold by a seller of honesty $h$. According to definition \req{gaso}, 
the aggregate satisfaction arising from \req{gammashm} reads:
\be \label{gammasm}
\Gamma_S(t)=\frac{B(t)}{S(t)}\left[(1+m)\avg{\acca}-\avg{h^{\alpha+1}} 
\right].
\ee
It is worth noticing the strong feedback effect contained in it: if $S(t)$ 
becomes much larger than $B(t)$, then $\Gamma_S(t)$ diminishes, thus slowing 
down the growing rate of $S(t)$ itself. As a consequence a stationary state 
is reached when $B(t)$ and $S(t)$ grow exponentially with the same exponent, 
which is entirely determined by $\lim_{t\rightarrow 
\infty}\Gamma_B(t)=\Gamma_B$. An example is given in figure \ref{fig2}.

In the limit of large $S$ we can employ the following approximation:
\be\label{appro}
\int_0^{q_{max}} \mu(q) dq \simeq 1-\frac{1}{S(t)+1},
\ee
where $\mu(q)$ is the overlap distribution. Equation \req{appro} becomes exact if $\mu(q)$ is uniform.
Let us assume, for the sake of simplicity, that products $x_{s}$ are real 
numbers uniformly distributed between zero and one. A natural definition of 
the distance between two products, on the torus $[0,1]$, is
\be\label{distance}
d_{b,s}=\min(|x_{b}-x_{s}|,1-|x_{b}-x_{s}|),
\ee
which yields the following overlap distribution:
\be
\mu(q)=2\Theta(q-0.5),
\ee
where $\Theta$ is the Heaviside function.
Equation \req{appro} then gives
\be\label{qm}
q_{max}=\frac{2S(t)+1}{2(S(t)+1)}.
\ee

\begin{figure}
\epsfig{file=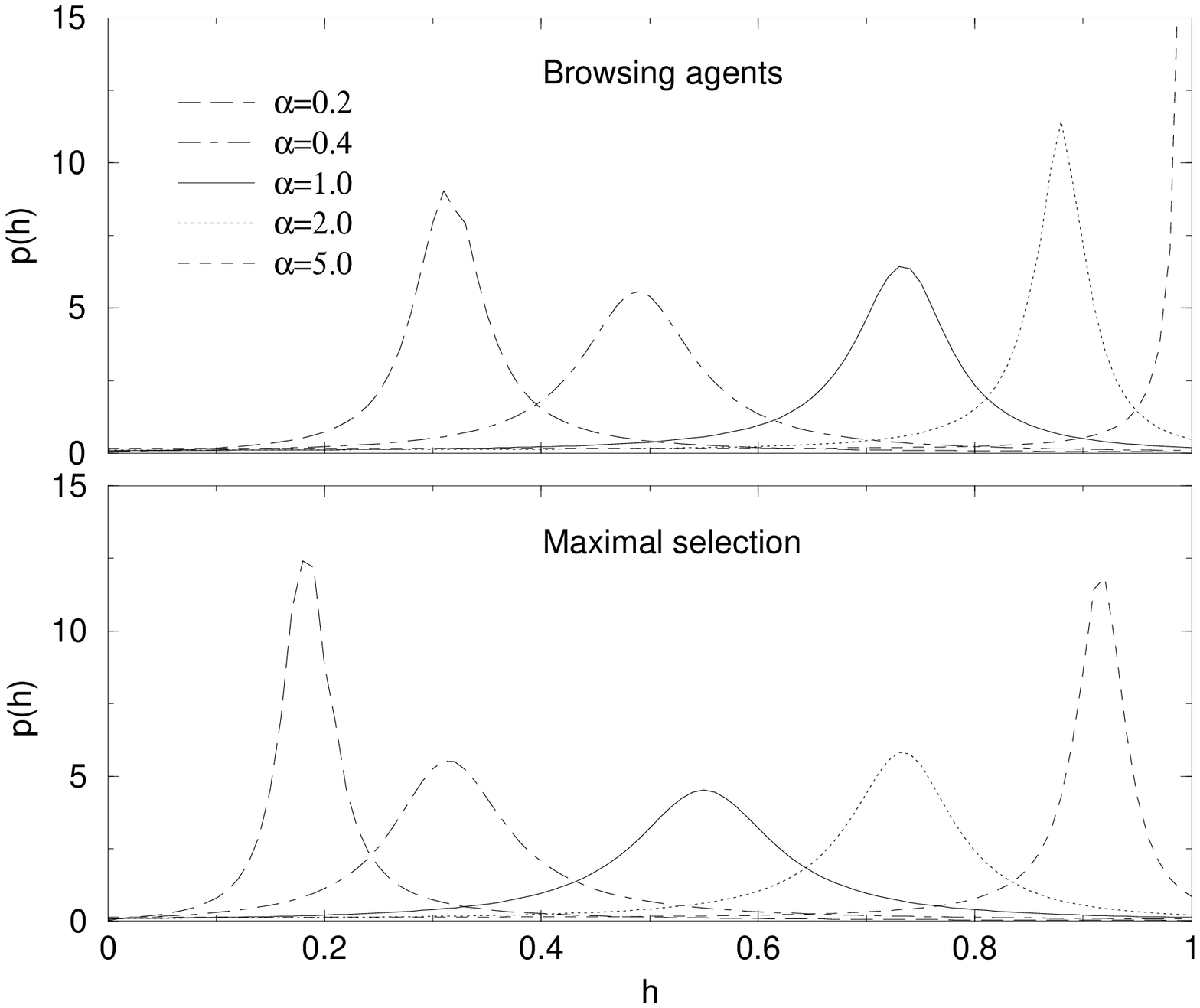, width=250pt}
\caption{Stationary honesty distribution of sellers with maximal selection 
(lower graph) and browsing agents with $\rho=0.5$ (upper graph). Different 
line-styles correspond to different values of $\alpha$: the legend refers to 
both graphs. We fixed $H=100$, $m=0.1$ and $c_B=0.9$. Normalization of 
$p(h)$ is set to $100$.} \label{dune2}
\end{figure}

We solved numerically equations \req{buydyn} and \req{seldyn}, with 
definition \req{gammabm} for buyers' satisfaction and definition 
\req{gammashm}, in the approximation \req{qm}, for that of sellers. 
Positing a uniform distribution at time zero, we focused on the honesty distribution of sellers in the stationary regime 
$p(h)=\lim_{t \to \infty}p(h,t)$. 
Since $p(h,t)$ \req{pidiac} results from a natural selection of sellers as a 
consequence of buyers' behavior, honesties appearing with greater 
probability reflect higher earnings realized by the corresponding sellers.
The lower graph of figure \ref{dune2} shows a shift of distribution $p(h)$ 
towards a greater average honesty $\avg{h}$, as the value of $\alpha$ is 
increased. In our model $\alpha$ is the relevant parameter: the larger it 
is, the more buyers take sellers' reputation into account. In fact the 
probability $f_b({\hat s_b})$ \req{pur} that buyer $b$ actually purchases 
product $x_{\hat s_b}$, decreases for greater $\alpha$. Such a decrease is 
not uniform in $h$, but scales as a power law. As a result, with increasing 
$\alpha$ sellers with higher honesty are more favored, their relative 
frequency is enhanced and so is buyer's probability of purchase.
The net result of these two competing effects is a greater buyers' 
satisfaction, in the stationary state, when $\alpha$ is bigger. This appears clearly in figure 
\ref{gamon}, where the average honesty $\avg{h}$ (upper graph) and the 
buyers' satisfaction $\Gamma_B$ (lower graph) are shown to be increasing 
functions of $\alpha$.
As already mentioned, $\Gamma_B$ determines the slope of both buyers and 
sellers exponential growth.
We conclude that a greater $\alpha$ exerts more selective pressure on 
sellers, giving rise to a more efficient market and to a faster growth of 
the web-site usage.

\section{Browsing agents}
If, instead of considering only the product that maximizes his overlap, a 
buyer also looks at other offers, he might find better deals.
To make things clear, imagine a parameter $\rho\in(0,0.5]$ tunes the width 
of customers' search for goods.
Among the $S$ items available in the market, buyer $b$ examines the ones 
($2S(t)\rho$ on average) closer than $\rho$ to his desired one, i.e. those 
that fulfill the condition $d_{b,s}<\rho$.
This mimics a situation where buyers browse the portion of the web-site 
containing products they might be interested in. This task is made easy by 
the division of products into categories, provided by most portals, and by 
the possibility to display first the ones sold by more reputable sellers.
Buyer $b$ can thus operate a quick selection, after which he picks only one 
product $s$, with probability $z_b(s)$, and analyzes it more closely.
In the preceding section we analyzed the case $\rho\rightarrow 0$, where 
$z_b(s)$ becomes a Dirac delta function centered in $x_{\hat s_b}$. We want 
to approach here the opposite limit, that of agents performing a wide search 
before evaluating something for purchase.

Once he has chosen an item $x_{\tilde s}$, buyer $b$ proceeds as before: he 
purchases it with probability $f_b(\tilde{s})$ \req{pur}, and is eventually 
rewarded with $r_{b,\tilde{s}}$ \req{rew}.
The average buyers' satisfaction over all transactions taking place at time 
$t$, namely $\Gamma_B(t)$, then reads
\be\label{gaco}
\Gamma_B(t)=\frac{1}{B(t)}\sum_b \sum_{s: d_{b,s}<\rho} z_b(s) f_b(s) 
r_{b,s}.
\ee

It is sensible to define $z_b(s)$ as a monotonically increasing function of 
$f_b(s)$. This means the probability of choosing a certain product for 
evaluation, is proportional to the probability of actually buying it 
afterwards. This is justified as long as items in the web-site are well 
organized and sorted. In order to be consistent with such an assumption, the 
exact functional form of $z_b(s)$ must somehow compensate the density of 
products available within a given portion of the space. If products are real 
numbers uniformly distributed in the domain $[0,1]$ and we adopt definition 
\req{distance} for the distance, then $\mu(q)$ is flat and we can simply set 
a linear dependence:
\be\label{cij}
z_b(s)=\frac{f_b(s)}{\sum_{s: d_{b,s}<\rho}f_b(s)}.
\ee
Let us define the conditional probability $\mu(q_{b,s}|x_{b})$ that a buyer 
$b$ has overlap $q_{b,s}$ with seller $s$, given his desire $x_{b}$. 
Inserting equation \req{cij} in \req{gaco}, and  employing definitions 
\req{rew} and \req{pur}, we obtain:
\bea\nonumber
\Gamma_B(t)&=&\frac{1}{B(t)}\sum_b \frac{\sum_{s: d_{b,s}<\rho} 
h^{2\alpha+1}_s 
q_{b,s}^3 }{\sum_{s: d_{b,s}<\rho} \acca_s q_{b,s}}\\\label{buysatint}
&\rightarrow& \frac{ \avg{h^{2\alpha+1}} }{ \avg{\acca} }\int_0^1 dx \frac{ 
\int_0^1 dq \mu(q|x) q^3 \Theta(q-\tilde{\rho}) }{ \int_0^1 dq \mu(q|x) q 
\Theta(q-\tilde{\rho}) },
\eea
where $\tilde{\rho}=1-\rho$ and the arrow stands for the limit of large $S$ 
and $B$, and for $\rho\gg 1/S$.
Similarly we can compute the average profit made by a seller belonging to a 
certain honesty level $h$:
\bea\nonumber
\Gamma_{S_h}(t)&=&\frac{g_h}{S_h(t)} \sum_b \frac{\sum_{s: [d_{b,s}<\rho 
\bigcap  h_s=h]} h^{2\alpha}_s q_{b,s}^2}{\sum_{s: d_{b,s}<\rho} \acca_s 
q_{b,s}}\\\label{Nvhint}
&\rightarrow& g_h \frac{B(t)}{S(t)} \frac{ h^{2\alpha} }{ \avg{h^{\alpha}} } 
\int_0^1 dx \frac{ \int_0^1 dq \mu(q|x) q^2 \Theta(q-\tilde{\rho})}{ 
\int_0^1 dq \mu(q|x) q \Theta(q-\tilde{\rho}) },
\eea
where $g_h$ is defined in \req{gain}. Here the limit is taken as in 
\req{buysatint}, with the additional condition $\rho \gg 1/S_h$ for every 
$h$.

It is easy to compute the conditional probability $\mu(q_{b,s}|x_{b})$. With 
definition \req{distance} of the distance we obtain, in the continuous 
limit:
\beas
\mu(q|x)=\int_0^1 dy \delta(q-\max[|x-y|,1-|x-y|])
=2\Theta(q-0.5).
\eeas
Equations \req{buysatint} and \req{Nvhint} become:
\bea\label{gammab}
\Gamma_B(t)&=&\frac{1+\tilde{\rho}^2}{2}\frac{\avg{h^{2\alpha+1}}}{\avg{\acca}}\\ 
\label{gammash}
\Gamma_{S_h}(t)&=&\frac{1-\tilde{\rho}^3}{1-\tilde{\rho}^2}\frac{2 B(t)}{3 
S(t)}\frac{h^{2\alpha}}{\avg{\acca}}(1-h+m)\\ \label{gammas}
\Gamma_{S}(t)&=&\frac{1-\tilde{\rho}^3}{1-\tilde{\rho}^2}\frac{2 B(t)}{3 
S(t)} \frac{1}{\avg{\acca}}
\left[(1+m)\avg{h^{2\alpha}}-\avg{h^{2\alpha+1}}\right].
\eea

\begin{figure}
\epsfig{file=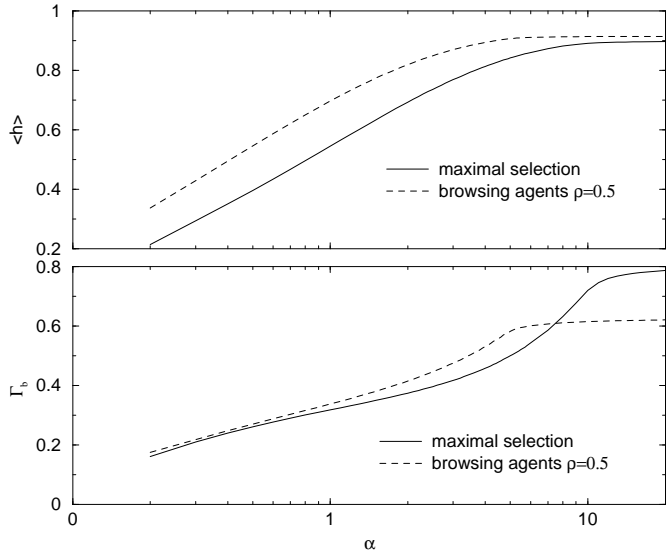, width=250pt}
\caption{Upper graph: average honesty of sellers, in the stationary state, 
as a function of $\alpha$. Lower graph: average buyers' satisfaction 
$\Gamma_B$, in the stationary state, as a function of $\alpha$. We fixed 
$H=100$, $m=0.1$ and $c_B=0.9$. The logarithmic x-axis scale is the same for 
both graphs.} \label{gamon}
\end{figure}

\begin{figure}
\epsfig{file=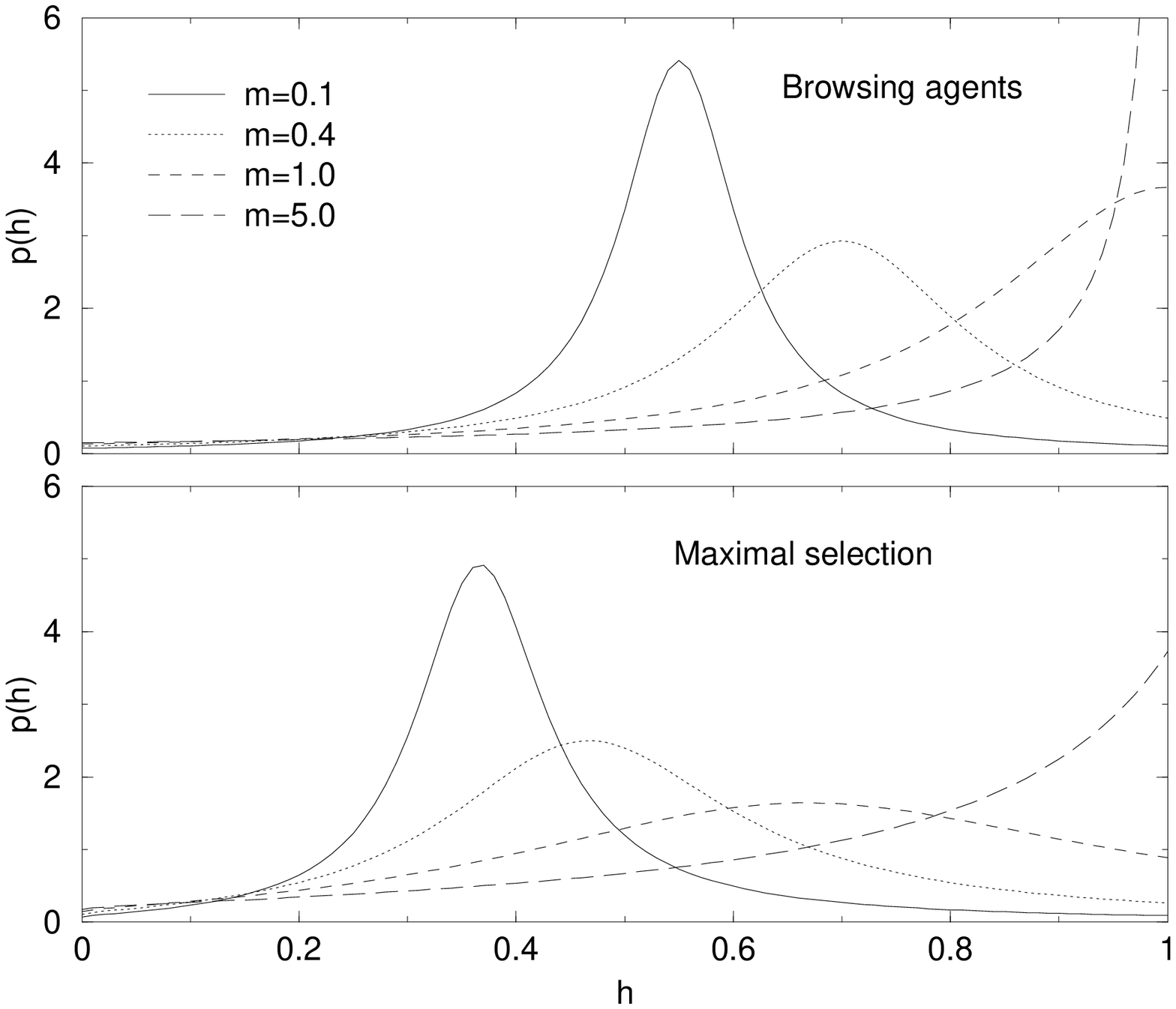, width=250pt}
\caption{Stationary honesty distribution of sellers with maximal selection 
(lower graph) and browsing agents with $\rho=0.5$ (upper graph). Different 
line-styles correspond to different values of the profit margin $m$: the 
legend refers to both graphs. We fixed $H=100$, $\alpha=0.5$ and $c_B=0.9$. 
Normalization of $p(h)$ is set to $100$.} \label{margra}
\end{figure}

\begin{figure}
\epsfig{file=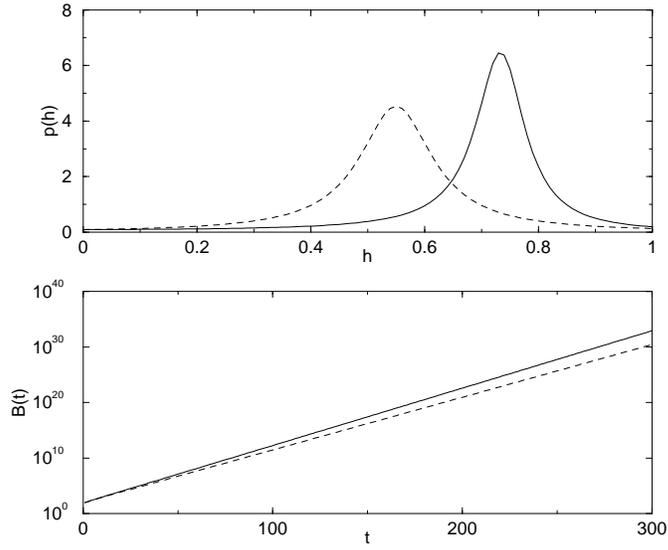, width=250pt}
\caption{Buyers' time evolution (lower graph) and stable honesty 
distribution of sellers (upper graph). Solid lines are browsing angents 
simulations with $\rho=0.5$, while dashed ones are with maximal selection. 
In both cases we fixed $H=100$, $m=0.1$ and $\alpha=1$.} \label{fig4}
\end{figure}

\begin{figure}
\epsfig{file=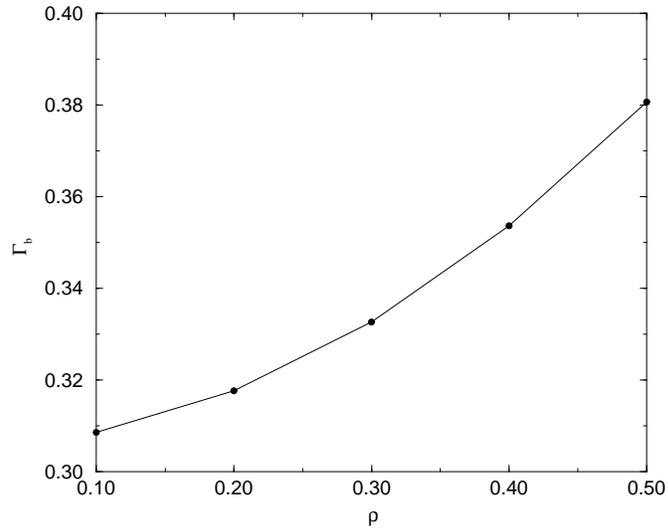, width=250pt}
\caption{Average buyers' satisfaction 
$\Gamma_B$ for browsing agents, in the stationary state, as a function of $\rho$. We fixed 
$H=100$, $m=0.1$, $c_B=0.9$ and $\alpha=1.5$.} \label{gammarho}
\end{figure}

We solved numerically equations \req{buydyn} and \req{seldyn} with the above 
definitions of the $\Gamma$-s and with a uniform initial honesty distribution of sellers.  
In the upper graphs of figures \ref{dune2} 
and \ref{margra} we show the $\alpha$ and $m$-dependence of the stable 
honesty distribution $p(h)$ for browsing agents. The lower 
graphs of these figures show, as a comparison, simulations with maximal 
selection. 
For any given set of the parameters, browsing buyers force sellers to be 
more honest than $q$-maximizing ones. This is also shown in the upper graph 
of figure \ref{gamon}, where the $\alpha$-dependence of average honesty is 
displayed. Now we can ask ourselves if also the web-site usage grows more 
with browsing agents than in the maximal selection case. In the lower graph 
of figure \ref{gamon} the stationary buyers' average satisfaction 
$\Gamma_B$, which governs the slope of the exponential growth of $B(t)$ and 
$S(t)$, is plotted against $\alpha$. Up to $\alpha\simeq7.5$ we certainly 
have a faster growth with browsing agents. A typical snapshot of this 
situation is given in figure \ref{fig4}, where the stationary honesty 
distribution and the time growth of $B(t)$ are shown in the two cases. 
For greater values of $\alpha$ the average honesty approaches a plateau, and so 
does $\Gamma_B$. This limit is rather unrealistic: the overlap $q$ plays 
nearly no role in the decision of purchase, being dominated by $\acca$. It 
becomes, therefore, more profitable to adopt the maximal selection strategy.
We should also stress that, in a competitive market, a higher average 
honesty of sellers would improve the overall web-site reputation, thus 
increasing the value of $c_B$ and, consequently, the growth rate of $B(t)$. 
We will, nevertheless, neglect this effect.
Finally, figure \ref{gammarho} shows that $\Gamma_B$ grows with $\rho$ ---and so does $\Gamma_S$. 
This confirms the Marriage Problem instance \cite{zhang}: 
increased information, even restricted to one side (in our case that of buyers), is beneficial to the whole society.

\section{Dynamical equilibrium}
It is useful to reformulate the dynamics, i.e. eqs. \req{buydyn}, 
\req{seldyn} and \req{seldyntot}, in terms of variables
\bea
\sigma(h,t)&=& p(h,t)/\Delta h\\
\eta(t)&=&B(t)/S(t),
\eea
whose time derivatives read:
\bea
\dot{\sigma}(h,t)&=&(\Gamma_S(t)-1)+\sigma(h,t)[\Gamma_{S_h}(t)-2\Gamma_S(t)+1]\\
\dot{\eta}(t)&=&\eta(t)[\Gamma_B(t)-2\Gamma_S(t)+1+c_B].
\eea
These variables eventually reach a constant value, due to the equilibration 
of two sets of competing effects. First, that of sellers' honesty: a greater 
$h$ level enhances the probability of selling a product (see \req{pur}), but 
reduces the unitary gain $g_h=1+m-h$ \req{gain}. Second, that of the ratio 
buyers/sellers: a bigger $\eta(t)$ means there are more buyers for each 
seller. This increases the average sellers' satisfaction $\Gamma_S(t)$, 
which in turn makes $S(t)$ increase, and $\eta(t)$ diminish.
The stationarity condition yields:
\bea\label{prima}
\Gamma_B+c_B-1=2(\Gamma_S-1)\\ \label{seconda}
\sigma(h)=\frac{(\Gamma_S-1)}{2\Gamma_S-\Gamma_{S_h}-1},
\eea
from which it is clear that the inequality $\Gamma_B>1-c_B$ must hold to 
ensure growth. Equation \req{seconda} is the result of our darwinian-type 
selection, which implies that the most frequent $h$-population be the most 
fit (satisfied). From equation \req{gammashm} (resp. \req{gammash}) we can 
compute the mode $h_m^{MS}$ (resp. $h_m^{BA}$) of distribution $p(h)$:
\bea\label{hmaxms}
h_m^{MS}&=&\frac{(1+m)\alpha}{1+\alpha}\\\label{hmaxba}
h_m^{BA}&=&\frac{2(1+m)\alpha}{1+2\alpha}.
\eea
When the mode equals one, fully honest sellers have an advantage over the 
others. If that happens for a given set of parameters $(m,\alpha)$, agents' 
satisfaction approaches a limit value. In figure \ref{gamon} this is shown, 
in particular, for the $\alpha$-dependence of $\Gamma_B$. Equations 
\req{hmaxms} and \req{hmaxba} explain why the plateau value is reached 
faster with browsing agents.

In order to find the stationary honesty distribution, we should solve
equation \req{prima} for $\eta$ and substitute the result into
\req{seconda}. For the case of browsing agents, by inserting
expressions \req{gammab}, \req{gammash} and \req{gammas} in  equations
\req{prima} and \req{seconda}, we obtain: \bea \label{eta}
\eta&=&\frac{(c_B+1)\avg{\acca}+a_1\avg{h^{2\alpha+1}}}{2a_2\avg{u_{\alpha}(h)}}\\
\label{sigma}
\sigma(h) &=& \left\{ 1+ \left[ 1-\frac{u_{\alpha}(h)}{\avg{u_{\alpha}(h)}} 
\right] v_{\alpha} \right\} ^{-1},
\eea
where $a_1=\frac{1+\tilde{\rho}^2}{2}$, $a_2=\frac{2(1-\tilde{\rho}^3)}{3(1-\tilde{\rho}^2)}$ and
\beas
u_{\alpha}(h)&=&h^{2\alpha}(1+m)-h^{2\alpha+1}\\
v_{\alpha}&=&\left[ 1- \frac{\avg{\acca}}{ a_2 \eta \avg{u_{\alpha}(h)} }  
\right]^{-1}=
1+\frac{2\avg{\acca}}{a_1 \avg{h^{2\alpha+1}}+(c_B-1)\avg{\acca}}.
\eeas
Now equation \req{sigma} can be solved self-consistently.

\begin{figure}
\epsfig{file=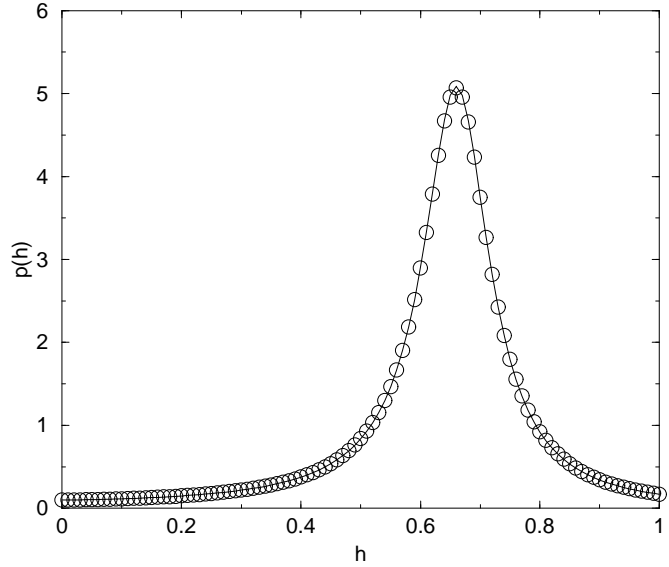, width=250pt}
\caption{Honesty distribution of sellers with maximal selection. Circles are 
numerical simulations, the solid line comes from equation \req{selfa}. The 
parameters are: $H=100$, $c_B=0.1$, $\alpha=1.5$ and $m=0.1$.} \label{coin}
\end{figure}

Similarly, for the case of maximal selection, we insert equations 
\req{gammabm}, \req{gammashm} and \req{gammasm} into \req{prima}, we 
eliminate $\eta$ and substitute the expressions thus obtained in 
\req{seconda}. Finally we end up with the following equation:
\be\label{selfa}
\left[ (\avg{h^{\alpha+1}}+c_B+1)\left( 1- \frac{{\tilde u}_{\alpha}(h)}{2 
\avg{{\tilde u}_{\alpha}(h)}} \right) 
-1\right]\sigma(h)=\frac{\avg{h^{\alpha+1}}+c_B-1}{2},
\ee
where ${\tilde u}_{\alpha}(h)$ is given by
\be\label{the}
{\tilde u}_{\alpha}(h)=h^{\alpha}(1+m)-h^{\alpha+1}.
\ee
The above relation \req{selfa} can be also solved self-consistently. An 
example is given in figure \ref{coin}, where the theoretical stationary 
distribution arising from \req{selfa} is shown to match exactly the one 
found solving numerically the original time dependent differential 
equations, \req{buydyn} and \req{seldyn}, with the same set of parameters. 
All other stationary quantities can be calculated accordingly.

\section{Honesty vs Reputation}
In the preceding sections we assumed reputation equals honesty. The two 
could, in fact, differ for the following main reasons. The first source of 
problem relies in imprecise consumers' ratings, but it is the minor one if 
the volume of affairs is big, since mistakes have no preferential direction. 
Moreover $\Delta h$ can be chosen of the same order of magnitude as the 
variance of individual mistakes, thus identifying $h$ with the consumers' 
average judgment. Second comes cheating, that is a seller, who has so far 
been good, might occasionally sell at an higher price. This could 
temporarily improve the gain of some seller, but it should only affect the 
variance and not the average satisfaction of buyers in the stationary 
regime. Thirdly, the reputation building process could be very inaccurate. 
We shall concentrate on the latter because it seems to be the main 
shortcoming of some commercial web-sites existing today.

Let us consider the extreme case, although common, where the rating form 
available in the web-site allows buyers to state if they made a good bargain 
or not, with no further specification. As a result, reputation consists in 
being good ($h_g$) or bad ($h_b$), and this is the only information about 
sellers buyers are provided with. Once they purchased a product, though, 
buyers can evaluate it accurately and judge it according to their proper 
honesty scale. Therefore the ``true'' honesty level $h$ still plays the same 
role here as in equation \req{rew}, whereas elsewhere it must be substituted 
by ${\tilde h}$, the two levels reputation. Equations \req{gammab}, 
\req{gammash} and \req{gammas} then become:
\bea\label{gammab2}
\Gamma_B(t)&=&\frac{1+\tilde{\rho}^2}{2}\frac{\avg{h {\tilde 
h}^{2\alpha}}}{\avg{{\tilde h}^{\alpha}}}\\ \label{gammash2}
\Gamma_{S_h}(t)&=&\frac{1-\tilde{\rho}^3}{1-\tilde{\rho}^2}\frac{2 B(t)}{3 
S(t)}\frac{{\tilde h}^{2\alpha}}{\avg{ {\tilde h}^{\alpha} }}(1-h+m)\\ 
\label{gammas2}
\Gamma_{S}(t)&=&\frac{1-\tilde{\rho}^3}{1-\tilde{\rho}^2}\frac{2 B(t)}{3 
S(t) \avg{ {\tilde h}^{\alpha} }}\left[(1+m)\avg{{\tilde 
h}^{2\alpha}}-\avg{h{\tilde h}^{2\alpha}}\right],
\eea
where ${\tilde h}=h_g$ if $h \ge 1/2$ and ${\tilde h}=h_b$ if $h < 1/2$.

\begin{figure}
\epsfig{file=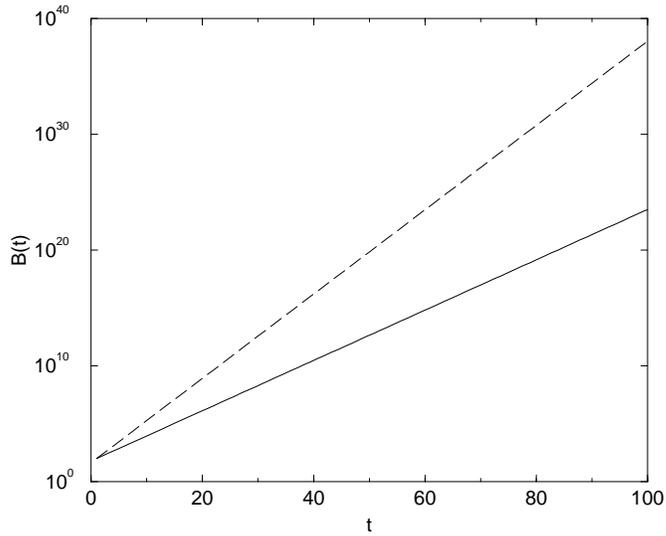, width=250pt}
\caption{Buyers growth as a function of time for browsing agents. The dashed 
line represent the case ${\tilde h=h}$, the solid line the binary case 
${\tilde h=h_g,h_b}$. The parameters are: $H=100$, $c_B=1.5$, $\alpha=1$ and 
$m=0.1$, in both cases.} \label{fig3}
\end{figure}

In this situation sellers less honest than $0.5$ tend to die out. For the 
higher intrinsic honesty levels, those who are closer to $0.5$ are favored, 
and $p(h)$ decays exponentially toward $h=1$. This defect of information 
transmission, something like a narrow channel effect \cite{shannon}, damages 
severely the web-site usage. In fig. \ref{fig3} we plotted the time 
evolution of $B(t)$ in this binary case, with $h_g=1-\Delta h$ and 
$h_b=\Delta h$, together with the case of browsing agents with perfect 
judging forms at their disposal. It is clear that the latter case shows a 
much faster growth.

\section{Comments}
We have shown, within our model, that a good rating form can help the growth 
of a commercial web-site, overcoming the problem of asymmetrical information. 
But, how is buyers' browsing ability influenced by 
its architecture? A good categorization of products is, of course, 
important: this way we would probably approach the most profitable region of 
figure \ref{gamon}.
We believe a major step forward would be achieved once it will be possible 
to guess accurately buyers' future wishes \cite{maslov,ykuo}.

The equations we studied, i.e. \req{buydyn} and \req{seldyn}, can be 
regarded as mean-field approximations to a stochastic behavior. We suppose, 
on average, an exponential growth of the web-site usage: this might mimic a 
fast growing stage of e-commerce web-sites. We believe the role of honesty 
and information we tried to stylize here applies to any situation where a 
great number of sellers is easily reachable to any buyer.

Our calculations are carried out by identifying buyers and sellers with real 
numbers: this is a useful simplification, but it is easy to substitute them 
with bit strings. In this case the distance \req{distance} becomes the 
hemming distance, and probability \req{cij} should be redefined 
appropriately.

Full information and unlimited processing capability of buyers could, in 
principle, allow them to maximize directly the product $hq$. Let us imagine 
each buyer $b$ follows the maximal selection strategy, with ${\hat 
s_b}=[{\hat s}: h_{\hat s} q_{b,{\hat s}}=\max_{s} h_s q_{b,s}]$. This would 
favor so much honest sellers that the honesty distribution $p(h)$ would 
become a delta function centered in $h=1$, which corresponds to a perfectly 
efficient market.

\section{Acknowledgements}
P.L. thanks A. Capocci and J. Waekeling for useful discussions. This work was supported by the Swiss National Fund, Grant No. 20-61470.00, and by the Grant Agency of the Czech Republic, Grant No. 202/01/1091.

\newpage

\end{document}